\documentstyle[aps,pra,preprint,floats,epsfig]{revtex}
\tightenlines
\begin{document}
\hfuzz 2pt \vfuzz 2pt \draft
\def\Journal#1#2#3#4{{\em #1}, {\bf #2}, #3 (#4)}
\title{\Large\bf Polarization-squeezed light formation in a medium with
electronic Kerr nonlinearity}
\author{F. Popescu\footnote{E-mail:\quad
florentin\_p@hotmail.com}}
\address{Physics Department, Florida State University, Tallahassee,
Florida, 32306}
\date{January 12, 2004}
\maketitle
\begin{abstract}
{\small {}\quad We analyze the formation of polarization-squeezed
light in a medium with electronic Kerr nonlinearity. Quantum
Stokes parameters are considered and the spectra of their quantum
fluctuations are investigated. It is established that the
frequency at which the suppression of quantum fluctuations is the
greatest can be controlled by adjusting the linear phase
difference between pulses. We shown that by varying the intensity
or the nonlinear phase shift per photon for one pulse, one can
effectively control the suppression of quantum fluctuations of the
quantum Stokes parameters.\\ PACS: 42.50.Dv, 42.50.Lc.}
\end{abstract}
\vspace{4mm}

{\bf Keywords}: {\it ultrashort light pulse, self- and cross-phase
modulation, electronic Kerr nonlinearity, quantum Stokes
parameters.}
\section{Introduction}
Knowledge of the quantum properties of the polarization of light
is essential when treating issues concerning
Einstein-Podolski-Rosen (EPR) paradox \cite{Ou} or Bell's
inequality \cite{Aspect}. The quantum analysis of the light
polarization is generally based on Stokes parameters associated
with Hermitian Stokes operators \cite{Agarwal,Tanas}. We call a
quantum state to be a polarization-squeezed (PS) state if it has
the level of quantum fluctuations of one of the quantum Stokes
parameters smaller than the level corresponding to the coherent
state \cite{Orlov}. Such a state can be generated by using the
optical parametric amplification ($\chi^{2}$)
\cite{Grangier,Bowen}. Recently, effective methods of PS state
formation using Kerr-like media ($\chi^{3}$) and optical solitons
in fibers have been proposed
\cite{Orlov,Silb,Korolkova,Heersnik,Oliver}. In the Kerr medium
the self-phase modulation (SPM) occurs, leading to quadrature
squeezing with preservation of the photon statistics
\cite{Kitagawa}.

It was noted in \cite{Blow} that a quantum treatment of SPM of
ultrashort light pulses (USPs) must account for the additional
noise related to the nonlinear absorbtion. The study of the
nonlinear propagation in Raman active media \cite{Boivin} accounts
for the quantum and thermal noises as a fluctuating addition to
the relaxation nonlinearity in the interaction Hamiltonian.
However, if we deal with USPs propagation, for instance through
fused-silica fibers, the electronic motion on $\sim 1$ fs time
scale contributes with about $80$\% to the Kerr effect while Raman
oscillators give only $20$\% contribution \cite{Shapiro}. An
attempt to develop the quantum theory of pulse SPM has been
undertaken in \cite{Boivin2}, where the electronic Kerr
nonlinearity is modelled as a Raman-like one. The electronic
nonlinearity was considered correctly with the interaction
Hamiltonian \cite{POP99} and the momentum operator \cite{POP00} in
the normally ordered form.

If two-mode radiation with orthogonal polarization and/or
different frequencies propagates through a Kerr medium, then the
cross-phase modulation (XPM) and parametric interaction can also
occur. The approach in \cite{POP99,POP00} has been extended for
the combined SPM--XPM of USPs in \cite{QSO}, by considering the
spectra of fluctuations of quadratures. Notice that, for
interacting solitons in a Kerr medium with anomalous dispersion
($k^{''}<0$), the XPM induces the transient photon-number
correlations while SPM answers for the intrapulse ones
\cite{Konig}.

The experimental realization of the polarization squeezing implies
the spatial overlapping of an orthogonally polarized strong
coherent beam with squeezed vacuum on a 50~:~50 beamsplitter
\cite{Hald} or the interference of two independent
quadrature-squeezed USPs produced in a fiber Sagnac interferometer
\cite{Silb}. The last series of experiments \cite{Heersnik,Oliver}
involve both the temporal and spatial overlaps of two orthogonally
polarized quadrature-squeezed USPs. From the theoretical point of
view, the quantum fluctuations of Stokes parameters in the case of
the nonlinear propagation in an isotropic Kerr medium were
introduced by the study of the light polarization \cite{Tanas}.
The time-independent quantum treatment of two-mode interaction
\cite{Orlov} predicted the possibility of forming the PS state in
anisotropic Kerr media.

In this letter we report on the theoretical computation of the
spectra of fluctuations of the Stokes parameters in the case of
two orthogonally polarized propagating USPs in an anisotropic
electronic Kerr medium. We employ simultaneously the SPM for
quadrature squeezing and the XPM for the PS state formation and
its control. The estimations are done in the frame of the quantum
theory for SPM-XPM of USPs developed in \cite{QSO}. The response
time of the electronic Kerr nonlinearity is accounted and the
dispersion of linear properties is described in the first
approximation of the dispersion theory. We begin with a brief
description of the model introduced in \cite{QSO} and based on the
momentum operators for the pulse fields. For consistency, we
present some elements of the algebra of time-dependent Bose
operators. By using such algebra we get the average values, the
correlation functions, and the spectra of quantum fluctuations of
Stokes parameters. Our concluding remarks close the paper.

\section{Quantum theory of combined SPM-XPM effect in electronic
Kerr medium} The quantum theory developed in \cite{QSO} is based on the
following momentum operators:
\begin{eqnarray}
\hat{G}^{(j)}_{\rm{spm}}(z)&=&\hbar\beta_{j}\int_{-\infty}^{\infty}dt
\int_{-\infty}^{t}H(t-t_{1})\hat{\mathbf{N}}\bigl
[\hat{n}_{j}(t,z)\hat{n}_{j}(t_{1},z)\bigl]dt_{1},\label{spm2}\\
\hat{G}_{\rm{xpm}}(z)&=&\hbar\tilde{\beta}\int_{-\infty}^{\infty}dt
\int_{-\infty}^{t}H(t-t_{1})\bigl[\hat{n}_{1}(t,z)\hat{n}_{2}(t_{1},z)
+\hat{n}_{1}(t_{1},z)\hat{n}_{2}(t,z)\bigl]\,dt_{1},\label{xpm2}
\end{eqnarray}
where $j$ is the pulse number, $\beta_{j}$ and $\tilde{\beta}$ are
nonlinear coefficients
($\beta_{j}=\hbar\omega_{0}k_{0}n_{2(j)}/8n_{0}V$
\cite{QSO,Ahmanov}) ($j=1,2$), $\hat{\mathbf{N}}$ is the normal
ordering operator, $H(t)$ is the nonlinear response function
[$H(t)\neq 0$ at $t\geq 0$ and $H(t)=0$ at $t<0$],
$\hat{n}_{j}(t,z)=\hat{A}^{+}_{j}(t,z)\hat{A}_{j}(t,z)$ is the
``photon number density" operator in the cross-section $z$ of the
medium, and $\hat{A}_{j}(t,z)$ and $\hat{A}^{+}_{j}(t,z)$ are the
photon anihilation and creation Bose operators with the
commutation relations $[\hat{A}_{j}(t_{1},z),
\hat{A}^{+}_{k}(t_{2},z)]=\delta_{jk}\delta{(t_{2}-t_{1})}$. The
thermal noise is neglected and the expressions (\ref{spm2}) and
(\ref{xpm2}) are averaged over the thermal fluctuations. We assume
that the Kerr effect is produced by the electronic motion with
characteristic time $\tau_{r}\leq 1$ fs. In our approach the pulse
duration $\tau_{p}$ is much greater than the relaxation time
$\tau_{r}$, and the Kerr medium is lossless and dispersionless;
that is, the pulses frequencies are off resonance.

In general, the induced third-order polarization $P^{(3)}$ of the
medium at the frequency $\omega$ is:
$P^{(3)}_{i}(\omega)=\sum_{j}[\chi^{(3)}_{1122}E_{i}(\omega)E_{j}(\omega)E^{*}_{j}(\omega)+
\chi^{(3)}_{1212}E_{j}(\omega)E_{i}(\omega)E^{*}_{j}(\omega)+
\chi^{(3)}_{1221}E_{j}(\omega)E_{j}(\omega)E^{*}_{i}(\omega)]$
\cite{Orlov,Shen}. Since the first two terms in $P^{(3)}$ involve
operator products accounted in (\ref{spm2}) and (\ref{xpm2}), the
last term introduces products of the form
$\hat{A}^{+^{2}}_{1}\hat{A}^{{}^{2}}_{2}+\hat{A}^{+^{2}}_{2}\hat{A}^{{}^{2}}_{1}$
which are neglected since they are connected with the parametrical
interaction. The conditions (also assumed here) under which such
terms can be neglected were discussed in \cite{Orlov}, where it
was noted that the parametric interaction can be neglected if
$\Delta z\gg1$ and the phase mismatch $\Delta=2(k_{2}-k_{1})$ is
large, namely
$|\Delta|>\rm{min}\{\tilde{\beta}|A_{1}(0)|,\tilde{\beta}|A_{2}(0)|\}$.
Here
$\tilde{\beta}=8\pi\omega^{2}(\chi^{(3)}_{1122}+\chi^{(3)}_{1212}+\chi^{(3)}_{1221})/c^{2}(k_{1}+k_{2})$,
and $k_{j}=(\omega/c)n_{j}$. Note that for circularly polarized
modes the last term in $P^{(3)}$ is absent, which is valid in
isotropic Kerr media where modes obey only SPM and XPM. For
elliptically polarized fields the last term in $P^{(3)}$ induces
the circular birefringence $\Delta
n_{c}=-(2\pi/n_{0})2\chi^{(3)}_{1221}(|E_{+}(\omega)|^{2}-|E_{-}(\omega)|^{2})$
\cite{Shen}. However, for the successful experimental overlapping
of the two quadrature-squeezed USPs a birefringence compensator is
required \cite{Heersnik,Oliver}. In general, $\Delta n_{c}$ can be
neglected when USPs propagate along some symmetry directions
\cite{Shen}.

Since in our case the Kerr effect is of an electronic origin, in
the absence of one- and two-photon and Raman resonances, $H(t)$
can be chosen as $H(t)=(1/\tau_r)\exp{\left(-t/\tau_r\right)}$ at
$t\geq 0$ \cite{Ahmanov}. Then with (\ref{spm2}) and (\ref{xpm2})
the space-evolution equation for $\hat{A}_{1}(t,z)$ in the moving
coordinate system ($z=z^{'}$, $t=t^{'}-z/u$, where $t^{'}$ is the
running time and $u$ is the group velocity),
\begin{equation}\label{part1}
\frac{\partial\hat{A}_{1}(t,z)}{\partial z}-\bigl[\hat{O}_{1}(t)
+\hat{\tilde{O}}_{2}(t)\bigl]\hat{A}_{1}(t,z)=0,
\end{equation}
has the solution given by
\begin{equation}\label{anih}
\hat{A}_{1}(t,z)=e^{\hat{O}_{1}(t)+\hat{\tilde{O}}_{2}(t)}\,\hat{A}_{0,1}(t).
\end{equation}
Here $\hat{O}_{j}(t)=i\gamma_{j}q[\hat{n}_{0,j}(t)]$,
$\hat{\tilde{O}}_{j}(t)=i\tilde{\gamma}q[\hat{n}_{0,j}(t)]$,
$\gamma_{j}=\beta_{j} z$, $\tilde{\gamma}=\tilde{\beta} z$,
$\hat{n}_{0,j}(t)=\hat{A}^{+}_{0,j}(t)\hat{A}_{0,j}(t)$ is the
``photon number density" operator at the entrance into the medium
$z=0$, and
$q[\hat{n}_{0,j}(t)]=\int_{-\infty}^{\infty}h(t_{1})\hat{n}_{0,j}(t-t_{1})dt_{1}$,
where $h(t)=H(|t|)$. $\hat{A}_{2}(t,z)$ can be derived by changing
the indexes $1\leftrightarrow 2$ in (\ref{anih}). In comparison
with the so-called nonlinear Schr\"{o}dinger equation, used in the
quantum theory of optical solitons (see \cite{Konig} and refs.
therein), in (\ref{part1}) the pulse dispersion spreading in the
medium is absent. This approach corresponds to the first
approximation of the dispersion theory \cite{Ahmanov}. Note that
the structure of $q[\hat{n}_{0,j}(t)]$ is like that of the linear
response in the quantum description of a USP spreading in the
second-order approximation.

The statistical features of pulses at the output of the medium can
be evaluated by using the algebra of time-dependent Bose operators
\cite{POP99,POP00,QSO}. In such algebra we have
\begin{eqnarray}
\hat{A}_{0,j}(t_{1})e^{\hat{O}_{j}(t_{2})}=e^{\hat{O}_{j}(t_{2})
+{\mathcal{D}}_{j}(t_{2}-t_{1})}\hat{A}_{0,j}(t_{1}),\quad
\hat{A}_{0,j}(t_{1})e^{\hat{\tilde{O}}_{j}(t_{2})}=e^{\hat{\tilde{O}}_{j}(t_{2})
+{\tilde{\mathcal{D}}}(t_{2}-t_{1})}\hat{A}_{0,j}(t_{1}),\label{permut2}
\end{eqnarray}
where
${\mathcal{D}}_{j}(t_{2}-t_{1})=i\gamma_{j}\,h(t_{2}-t_{1})$,
${\tilde{\mathcal{D}}}(t_{2}-t_{1})=i\tilde{\gamma}h(t_{2}-t_{1})$,
$2\gamma_{j}$ is the nonlinear phase shift per photon for the
$j$-th pulse, and $\tilde{\gamma}$ is the nonlinear coupling
coefficient. By using the theorem of normal ordering
\cite{POP99,POP00}, one obtains the average values of the Bose
operators over the initial coherent summary state
$|\alpha_{0}(t)\rangle=|\alpha_{0,1}(t)\rangle\otimes|\alpha_{0,2}(t)\rangle$:
\begin{eqnarray}
\langle e^{\hat{O}_{j}(t)}\rangle &=& e^{i\phi_{j}(t)-\mu_{j}(t)},\quad
~\langle e^{\hat{\tilde{O}}_{j}(t)}\rangle
=e^{i\tilde{\phi}_{j}(t)-\tilde{\mu}_{j}(t)},\label{hat1}\\
\langle e^{\hat{O}_{j}(t_{1})+\hat{O}_{j}(t_{2})}\rangle &=&
e^{i[\phi_{j}(t_{1})+\phi_{j}(t_{2})]-[\mu_{j}(t_{1})+\mu_{j}(t_{2})]-
{\mathcal{K}}_{j}(t_{1},t_{2})},\label{susi1}\\ \langle
e^{\hat{\tilde{O}}_{j}(t_{1})+\hat{\tilde{O}}_{j}(t_{2})}\rangle &=&
e^{i[\tilde{\phi}_{j}(t_{1})+\tilde{\phi}_{j}(t_{2})]-
[\tilde{\mu}_{j}(t_{1})+\tilde{\mu}_{j}(t_{2})]-\tilde{\mathcal{K}}_{j}(t_{1},t_{2})}.\label{tutu1}
\end{eqnarray}
The parameters $\phi_{j}(t)=2\gamma_{j}\bar{n}_{0,j}(t)$,
$\mu_{j}(t)=\gamma^{2}_{j}\bar{n}_{0,j}(t)/2$ are connected with
SPM of the $j$-th pulse and
$\tilde{\phi}_{j}(t)=2\tilde{\gamma}\bar{n}_{0,j}(t)$,
$\tilde{\mu}_{j}(t)=\tilde{\gamma}^{2}\bar{n}_{0,j}(t)/2$ with the
pulses' XPM. Here $\phi_{j}(t)$ ($\tilde{\phi}_{j}(t)$) is the
nonlinear phase addition caused by SPM (XPM). The eigenvalue
$\alpha_{0,j}(t)$ of the annihilation operator $\hat{A}_{0,j}(t)$
over the coherent state $|\alpha_{0,j}(t)\rangle$ can be written
as $\alpha_{0,j}(t)=|\alpha_{0,j}(t)| e^{i\varphi_{j}(t)}$, where
$\varphi_{j}(t)$ is the linear phase of the $j$-th pulse. Then
$|\alpha_{0,j}(t)|^{2}=\langle\hat{n}_{0,j}(t)\rangle\equiv\bar{n}_{0,j}(t)$.
For simplicity let $\bar{n}_{0,j}(t=0)\equiv\bar{n}_{0,j}$. We
separate the time dependence in $|\alpha_{0,j}(t)|$ by introducing
the pulse's envelope $r_{j}(t)$ so that
$|\alpha_{0,j}(t)|=|\alpha_{0,j}(0)|r_{j}(t)$ with $r_{j}(0)=1$.
In (\ref{susi1})-(\ref{tutu1}),
${\mathcal{K}}_{j}(t_{1},t_{2})=\mu_{0,j}\,r^{2}_{j}(t_{1}+\tau/2)g(\tau)$
and
$\tilde{\mathcal{K}}_{j}(t_{1},t_{2})=\tilde{\mu}_{0,j}\,r^{2}_{j}(t_{1}+\tau/2)g(\tau)$
are the temporal correlators, where
$\mu_{0,j}=\gamma_{j}\bar{n}_{0,j}$,
$\tilde{\mu}_{0,j}=\tilde{\gamma}\bar{n}_{0,j}$,
$g(\tau)=(1+|\tau|/\tau_{r})h(\tau)$, and $\tau=t_{2}-t_{1}$.

\section{Quantum Stokes parameters and their average values, polarization degree}\label{section3}
In classical optics the polarization state is visualized as a
Stokes vector on the Poincar\'e sphere and is characterized by the
four classical Stokes parameters $\{S_{k}\}_{k=0,3}$. Since
$S_{0}$ defines the total intensity of pulse field,
$\{S_{k}\}_{k=1,3}$ characterize light polarization and form a
Cartesian axis system. Each point on the sphere corresponds to a
definite polarization state, whose variation is characterized by
the motion of the point on the sphere. In quantum optics,
$\{S_{k}\}_{k=0,3}$ are replaced by the operators
$\{\hat{S}_{k}\}_{k=0,3}$ which obey: $[\hat{S}_{0},
\hat{S}_{i}]=0$, $[\hat{S}_{i},
\hat{S}_{j}]=2i\varepsilon_{ijk}\hat{S}_{k}$. We define the
quantum Stokes parameters as:
$\hat{S}_{0}(t,z)=\sum^{2}_{j=1}\hat{n}_{j}(t,z)$,
$\hat{S}_{1}(t,z)=\hat{n}_{1}(t,z)-\hat{n}_{2}(t,z)$,
\begin{eqnarray}
\hat{S}_{2}(t,z)&=&\hat{A}^{+}_{2}(t,z)\hat{A}_{1}(t,z)+\hat{A}^{+}_{1}(t,z)\hat{A}_{2}(t,z),\label{S0}\\
\hat{S}_{3}(t,z)&=&i~[\hat{A}^{+}_{2}(t,z)\hat{A}_{1}(t,z)-\hat{A}^{+}_{1}(t,z)\hat{A}_{2}(t,z)],\label{S3}
\end{eqnarray}
where in our case $\hat{A}_{j}(t,z)$ is given by (\ref{anih}) and
$\hat{A}^{+}_{j}(t,z)$ is its Hermitian conjugate. By using Eqs.\
(\ref{permut2})-(\ref{hat1}) we compute the average values of
$\{\hat{S}_{k}(t,z)\}_{k=0,3}$ on the coherent summary state
$|\alpha_{0}(t)\rangle$. Finally, we get:
$\langle\hat{S}_{0}(t,z)\rangle=\sum^{2}_{j=1}\bar{n}_{0,j}(t)$,
$\langle\hat{S}_{1}(t,z)\rangle=\bar{n}_{0,1}(t)-\bar{n}_{0,2}(t)$,
and
\begin{eqnarray}
\langle\hat{S}_{2}(t,z)\rangle&=&2{[\bar{n}_{0,1}(t)\bar{n}_{0,2}(t)]}^{1/2}e^{-[\Delta_{1}(t)+\Delta_{2}(t)]}
\cos{[\tilde{\Phi}_{2}(t)-\tilde{\Phi}_{1}(t)]},\label{med2}\\
\langle\hat{S}_{3}(t,z)\rangle&=&2{[\bar{n}_{0,1}(t)\bar{n}_{0,2}(t)]}^{1/2}e^{-[\Delta_{1}(t)+\Delta_{2}(t)]}
\sin{[\tilde{\Phi}_{2}(t)-\tilde{\Phi}_{1}(t)]},\label{med3}
\end{eqnarray}
where $\Delta_{j}(t)=\mu_{j}(t)+\tilde{\mu}_{j}(t)$,
$\tilde{\Phi}_{j}(t)=\phi_{j}(t)-\tilde{\phi}_{j}(t)+\varphi_{j}(t)$.
By measurements, one obtains a set of measurable quantities
associated with the operators (\ref{S0}) and (\ref{S3}). However,
the presence of quantum fluctuations results in an uncertainty for
the measured quantities. The fluctuation uncertainty of each
$\langle\hat{S}_{k}(t,z)\rangle$ can be associated with a
particular region of uncertainty on the Poincar\'e sphere. In the
case of the nonlinear propagation, the ball-region of uncertainty,
specifically for the coherent state of light, changes to the
ellipsoid of uncertainty \cite{Orlov}.

In classical optics the polarization degree ${\mathcal{P}}$ is the
ratio of the intensity of the polarized part of the radiation
$I_{\text{pol}}$ to the total intensity $I_{\text{tot}}$, and is
connected with the classical Stokes parameters by
${\mathcal{P}}={\mathcal{R}}/\langle S_{0}\rangle$, where
${\mathcal{R}}$ is the radius of the classical Poincar\'e sphere,
${\mathcal{R}}^{2}=\sum_{k=1}^{3}\langle S_{k}\rangle^{2}$. Since
for completely polarized light ${\mathcal{P}}=1$, for partially
polarized light $0<{\mathcal{P}}<1$. The definition of the quantum
polarization degree follows the classical one:
${\mathcal{P}}(t,z)={{\mathcal{R}}(t,z)}/\langle\hat{S}_{0}(t,z)\rangle$,
where
${\mathcal{R}}^{2}(t,z)=\sum_{k=1}^{3}\langle\hat{S}_{k}(t,z)\rangle^{2}$.
With the expressions (\ref{med2})-(\ref{med3}) we have
\begin{equation}\label{arici}
{\mathcal{P}}(t,z)=\{1-4\bar{n}_{0,1}(t)\bar{n}_{0,2}(t)[\bar{n}_{0,1}(t)+\bar{n}_{0,2}(t)]^{-2}(1-
e^{-2[\Delta_{1}(t)+\Delta_{2}(t)]})\}^{1/2}.
\end{equation}
Since in real-life situations $\Delta_{1}(t)+\Delta_{2}(t)\ll1$,
we have ${\mathcal{P}}(t,z)\approx 1$. Indeed, a recent study
\cite{Agarwal,Orlov} investigated the nonlinear behavior of
${\mathcal{P}}(t,z)$ for the nonclassical states of light and
revealed that the deviation of ${\mathcal{P}}(t,z)$ from $1$ is a
pure quantum effect. Note that the results
(\ref{med2})-(\ref{arici}) are similar to the ones obtained in
\cite{Orlov}.

\section{Correlation functions and spectra of quantum Stokes parameters}\label{section4}
Since the Kerr medium is assumed to be lossless and
dispersionless, the operators $\hat{S}_{0}(t,z)$ and
$\hat{S}_{1}(t,z)$, as well as their dispersions, are conserved.
Therefore, we focus our attention to the quantum fluctuations of
$\hat{S}_{2}(t,z)$ and $\hat{S}_{3}(t,z)$ by defining their
correlation functions as
\begin{eqnarray}
R_{S_{k}}(t_{1},t_{2})=\left\langle\hat{S}_{k}(t_{1},z)\hat{S}_{k}(t_{2},z)\right\rangle-
\left\langle\hat{S}_{k}(t_{1},z)\right\rangle\left\langle\hat{S}_{k}(t_{2},z)\right\rangle\qquad
(k=2,3).\label{aud}
\end{eqnarray}
The correlation functions (\ref{aud}) can be analytically computed
by using Eqs.\ (\ref{permut2}) and (\ref{susi1})-({\ref{tutu1}}).
Here we simply write down the result for $R_{S_{2}}(t,t+\tau)$ in
the approximation $\gamma_{j},\tilde{\gamma}\ll 1$:
\begin{eqnarray}
R_{S_{2}}(t,t+\tau)&=&\delta(\tau)+
h(\tau)[\bar{n}_{0,1}(t)\phi_{2}(t)-\bar{n}_{0,2}(t)\phi_{1}(t)]
\sin{2[\tilde{\Phi}_{1}(t)-\tilde{\Phi}_{2}(t)]}\nonumber\\
&+&g(\tau)\{\bar{n}_{0,1}(t)[\phi^{2}_{2}(t)\!+\!\tilde{\phi}^{2}_{2}(t)]+
\bar{n}_{0,2}(t)[\phi^{2}_{1}(t)\!+\!\tilde{\phi}^{2}_{1}(t)]\}
\sin^{2}{[\tilde{\Phi}_{1}(t)\!-\!\tilde{\Phi}_{2}(t)]},\label{re1}
\end{eqnarray}
where for simplicity we denote $t_{1}=t$, $t_{2}=t+\tau$. In the
absence of SPM [$\phi_{j}(t)=0$] and XPM [$\tilde{\phi}_{j}(t)=0$]
of pulses, Eq.\ (\ref{re1}) gives us the correlation function for
the coherent state $R_{S_{2}}^{coh}(t,t+\tau)=\delta(\tau)$, as
expected. The spectral densities of quantum fluctuations of
$\hat{S}_{k}(t,z)$ can be evaluated by using Wiener-Khintchine
theorem:
$S_{S_{k}}(\omega,t)=\int_{-\infty}^{\infty}R_{S_{k}}(t,t+\tau)
e^{i\omega\tau}d\tau$. Allowing for a small change of the envelope
during the relaxation time we obtain
\begin{eqnarray}
S_{S_{2}}(\Omega,t)&=&1+2L(\Omega)[\bar{n}_{0,1}(t)\phi_{2}(t)-\bar{n}_{0,2}(t)\phi_{1}(t)]
\sin{2[\tilde{\Phi}_{1}(t)-\tilde{\Phi}_{2}(t)]}\nonumber\\
&+&4L^{2}(\Omega)\{\bar{n}_{0,1}(t)[\phi^{2}_{2}(t)\!+\!\tilde{\phi}^{2}_{2}(t)]+
\bar{n}_{0,2}(t)[\phi^{2}_{1}(t)\!+\!\tilde{\phi}^{2}_{1}(t)]\}
\sin^{2}{[\tilde{\Phi}_{1}(t)\!-\!\tilde{\Phi}_{2}(t)]},\label{S2Spectra}
\end{eqnarray}
where $L(\Omega)=1/(1+\Omega^{2})$, and $\Omega=\omega\tau_{r}$ is
the reduced frequency. The spectral density (\ref{S2Spectra})
depends on the relaxation time $\tau_{r}$ and quasi-statically
changes with time. Besides, the second term on r.h.s. of Eq.\
(\ref{S2Spectra}) indicates that the quantum fluctuations of
$\hat{S}_{2}(t,z)$ can be less than those corresponding to the
coherent state $S_{S_{2}}^{coh}(\Omega,t)=1$. The correlation
function of $\hat{S}_{3}(t,z)$, as well as its spectrum, can be
easily obtained by shifting (\ref{re1}) and (\ref{S2Spectra}) in
phase with $\pi/2$.

For an arbitrary $\Omega_{0}=\omega_{0}\tau_{r}$, at which the
linear phase difference between incoming pulses
\begin{eqnarray}\label{phasedif}
\Delta\varphi(t)_{\rm opt}
&=&\frac{1}{2}\arctan\left(\frac{\bar{n}_{0,1}(t)\phi_{2}(t)-\bar{n}_{0,2}(t)\phi_{1}(t)}
{L(\Omega_{0})\Bigl\{\bar{n}_{0,1}(t)[\phi^{2}_{2}(t)+\tilde{\phi}^{2}_{2}(t)]+
\bar{n}_{0,2}(t)[\phi^{2}_{1}(t)+\tilde{\phi}^{2}_{1}(t)]\Bigl\}}\right)\nonumber\\
&{}&+\phi_{1}(t)-\phi_{2}(t)-\tilde{\phi}_{1}(t)+\tilde{\phi}_{2}(t),
\end{eqnarray}
is optimized, the expression (\ref{S2Spectra}) reaches the minimum
value:
\begin{eqnarray}\label{S2min}
S_{S_{2}}(\Omega_{0},t)&=&1+2L^{2}(\Omega_{0})\{\bar{n}_{0,1}(t)[\phi^{2}_{2}(t)+\tilde{\phi}^{2}_{2}(t)]+
\bar{n}_{0,2}(t)[\phi^{2}_{1}(t)+\tilde{\phi}^{2}_{1}(t)]\}\nonumber\\
&{}&-2L(\Omega_{0})\Bigl([\bar{n}_{0,1}(t)\phi_{2}(t)-\bar{n}_{0,2}(t)\phi_{1}(t)]^{2}\nonumber\\
&{}&+L^{2}(\Omega_{0})\{\bar{n}_{0,1}(t)[\phi^{2}_{2}(t)+\tilde{\phi}^{2}_{2}(t)]+
\bar{n}_{0,2}(t)[\phi^{2}_{1}(t)+\tilde{\phi}^{2}_{1}(t)]\}^{2}\Bigl)^{1/2}.
\end{eqnarray}

\begin{figure}
\centering
\includegraphics[height=0.5\textwidth]{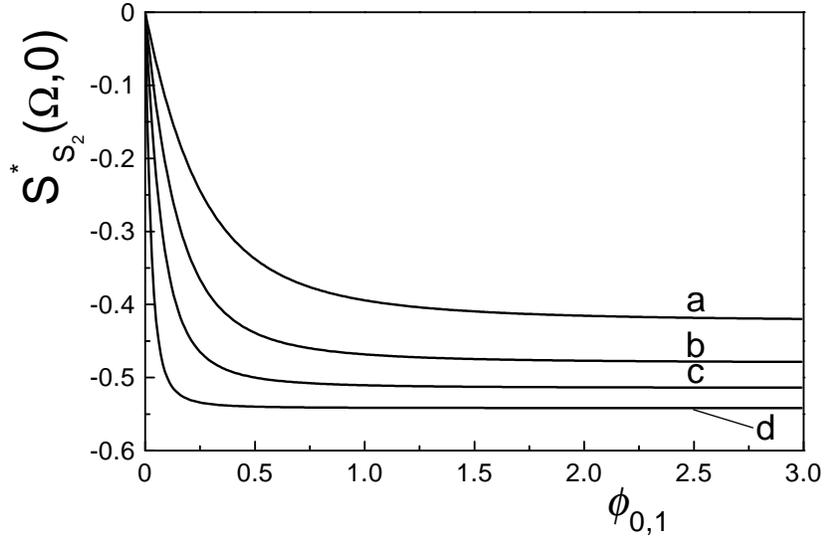}
\caption{Normalized spectral variance $S^{*}_{S_{2}}(\Omega,t)$ as
a function of the maximum nonlinear phase addition $\phi_{0,1}$ at
the reduced frequency $\Omega=0$ for initial phase difference
$\Delta\varphi(t)$ which is optimal at $\Omega_{0}=0$. Curves are
calculated at time $t=0$, $\gamma_1=\gamma_{2}/4=2\tilde{\gamma}$,
and correspond to $\bar{n}_{0,2}=\bar{n}_{0,1}/4$ (a),
$\bar{n}_{0,2}=\bar{n}_{0,1}/2$ (b), $\bar{n}_{0,2}=\bar{n}_{0,1}$
(c), $\bar{n}_{0,2}=3\bar{n}_{0,1}$ (d).} \label{Fig1}
\end{figure}

\begin{figure}
\centering
\includegraphics[height=0.5\textwidth]{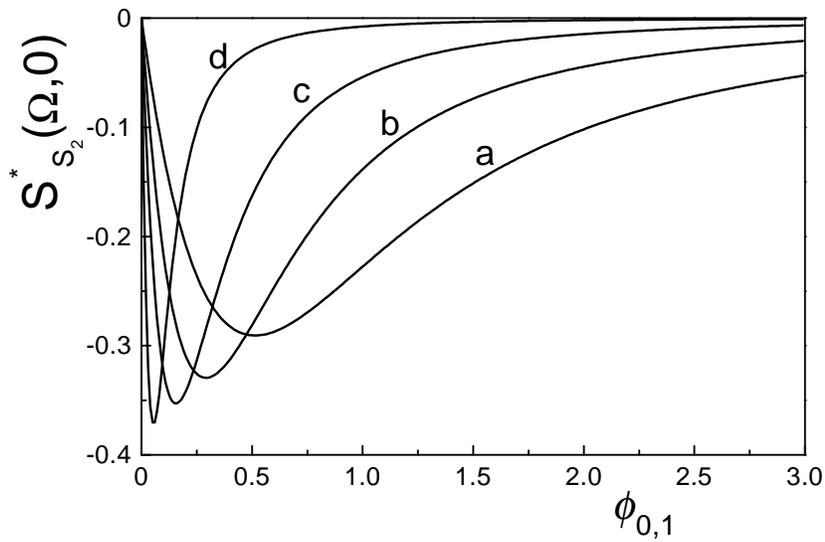}
\caption{As in Fig.\ \ref{Fig1} but for $\Omega_{0}=1$.}
\label{Fig2}
\end{figure}

\begin{figure}
\centering
\includegraphics[height=0.5\textwidth]{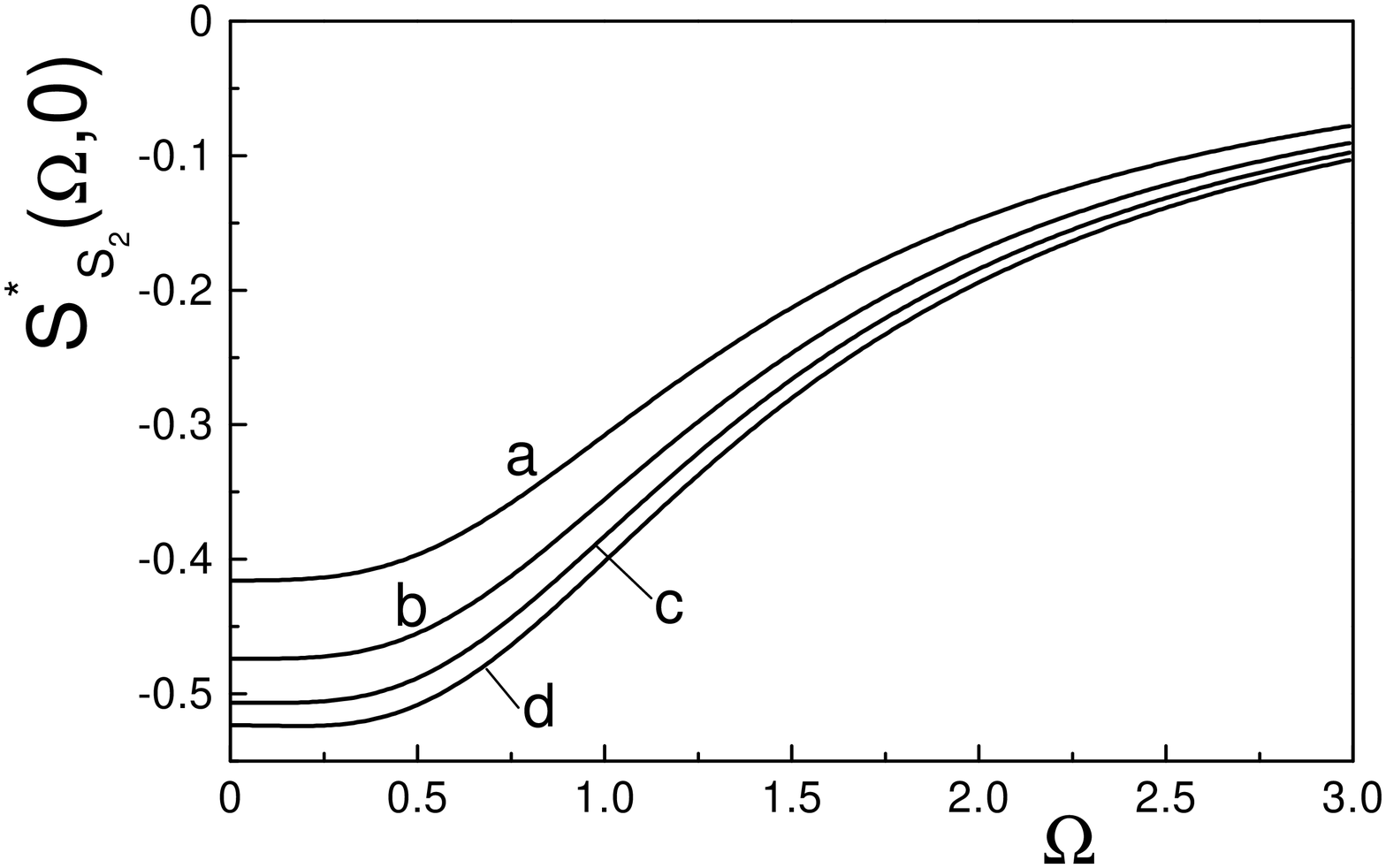}
\caption{Normalized spectral variance $S^{*}_{S_{2}}(\Omega,t)$ at
$\phi_{0,1}=2$ for initial phase difference $\Delta\varphi(t)$
 which is optimal at $\Omega_{0}=0$. Curves are calculated at time $t=0$,
$\gamma_1=\gamma_{2}/4=2\tilde{\gamma}$, and correspond to
$\bar{n}_{0,2}=\bar{n}_{0,1}/4$ (a),
$\bar{n}_{0,2}=\bar{n}_{0,1}/2$ (b), $\bar{n}_{0,2}=\bar{n}_{0,1}$
(c), $\bar{n}_{0,2}=3\bar{n}_{0,1}$ (d).} \label{Fig3}
\end{figure}

To characterize the deviation of $S_{S_{2}}(\Omega,t)$ from the
coherent level, we define the normalized spectral variance
$S^{*}_{S_{2}}(\Omega,t)=[S_{S_{2}}(\Omega,t)-1]/\bar{n}_{0,1}(t)$.
In the case of the suppression of quantum fluctuations $-1\leq
S^{*}_{S_{2}}(\Omega,t)<0$. Let us investigate the
$S^{*}_{S_{2}}(\Omega,t)$ by choosing the linear phase difference
between pulses to be optimal at a defined reduced frequency
$\Omega_{0}$, and change the intensity of one pulse (the control
pulse) in comparison with the intensity of the other one. Such
dependence of $S^{*}_{S_{2}}(\Omega,t)$ at $t=0$ and $\Omega=0$ on
the maximum nonlinear phase addition
$\phi_{0,1}\equiv\phi_{1}(t=0)$ in case the linear phase
difference is optimal at $\Omega_{0}=0$ and $\Omega_{0}=1$, is
displayed in Figs.\ \ref{Fig1} and \ref{Fig2}, respectively.
Thus, at $\Omega_{0}=0$ the increase of the control pulse
intensity produces a uniform suppression of quantum fluctuation of
$\hat{S}_{2}(t,z)$ for any $\phi_{0,1}>1$. At $\Omega_{0}=1$ the
increase of the control pulse intensity produces the suppression
basically in the domain $\phi_{0,1}<1$. The normalized spectral
variance $S^{*}_{S_{2}}(\Omega,t)$ at $t=0$, $\Omega_{0}=0$, and
fixed $\phi_{0,1}=2$, is presented in Fig.\ \ref{Fig3}.
Now the suppression in $\hat{S}_{2}(t,z)$ is maximal at
$\Omega\approx 0$. Summarizing, the choice of the linear phase
difference allows us to obtain the spectra with the form of
interest, and the increase of the control pulse intensity can
effectively control the suppression of quantum fluctuations of
quantum Stokes parameters.

\begin{figure}
\centering
\includegraphics[height=0.5\textwidth]{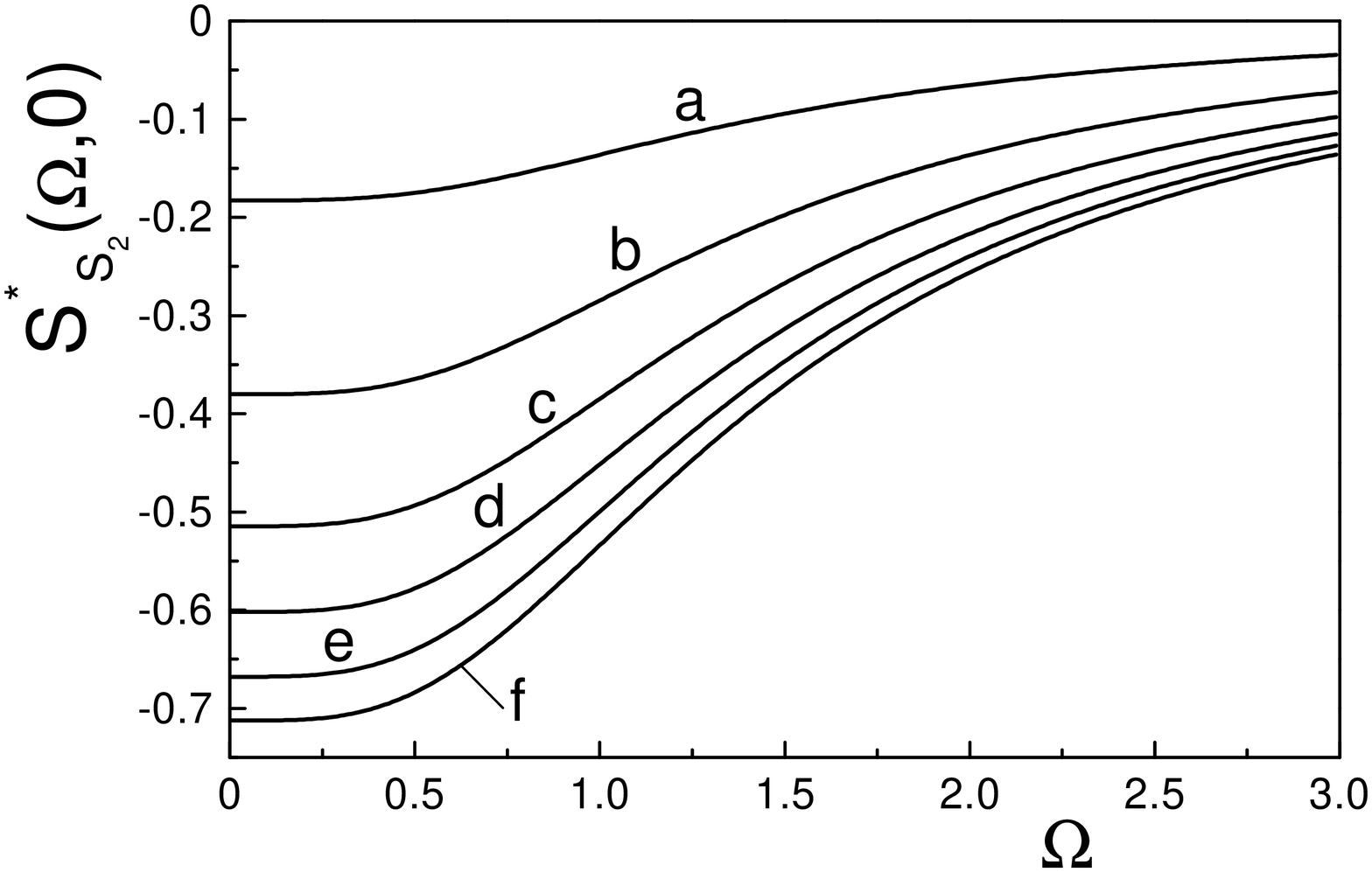}
\caption{Normalized spectral variance $S^{*}_{S_{2}}(\Omega,t)$ at
$\phi_{0,1}=2$ for initial phase difference $\Delta\varphi(t)$
which is optimal at $\Omega_{0}=0$. Curves are calculated at time
$t=0$, $\bar{n}_{0,1}=\bar{n}_{0,2}$,
$\tilde{\gamma}=\gamma_{1}/2$ and correspond to
$\gamma_{2}=2\gamma_{1}$ (a), $\gamma_{2}=3\gamma_1$ (b),
$\gamma_{2}=4\gamma_1$ (c), $\gamma_{2}=5\gamma_1$ (d),
$\gamma_{2}=6\gamma_1$ (e), $\gamma_{2}=7\gamma_1$ (f).}
\label{Fig4}
\end{figure}

\begin{figure}
\centering
\includegraphics[height=0.5\textwidth]{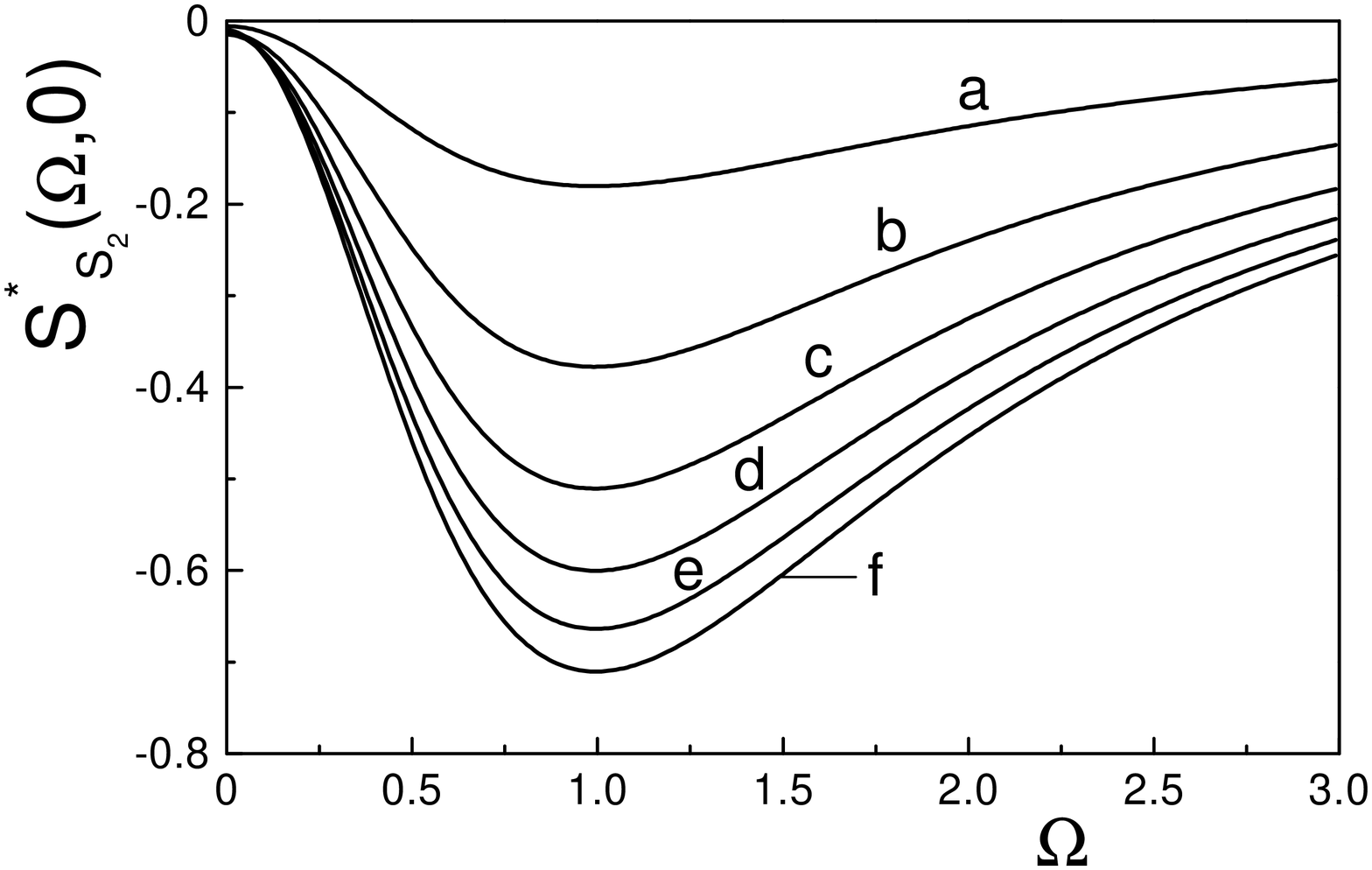}
\caption{As in Fig.\ \ref{Fig4} but for $\Omega_{0}=1$}
\label{Fig5}
\end{figure}

One can also control the suppression of quantum fluctuations of
$\hat{S}_{2}(t,z)$ by increasing the nonlinear coefficient
$\gamma_{2}$ in comparison with $\gamma_{1}$. This is equivalent
to the increase of the Kerr electronic nonlinearity for one pulse
$n_{2(2)}$ in comparison with the one for another pulse $n_{2(1)}$
($\gamma_{j}=\beta_{j}z$). The variance $S^{*}_{S_{2}}(\Omega,t)$
at $t=0$, $\Omega_{0}=0$ for various relations between nonlinear
coefficients $\gamma_{1}$ and $\gamma_{2}$ is displayed in Fig.\
\ref{Fig4} in the simplest case of pulses with the same intensity.
In this case, the increase of $\gamma_{2}$ in comparison with
$\gamma_{1}$ suppresses the quantum fluctuations of
$\hat{S}_{2}(t,z)$ basically at low frequencies, $\Omega\approx 0$
. A similar dependence, but at $\Omega_{0}=1$, is shown in Fig.\
\ref{Fig5}.
Now, the squeezing takes place essentially at $\Omega\approx 1$
($\omega\approx1/\tau_{r}$).

Note the experimentally obtained squeezing of $-2.8$ dB in
$\hat{S}_{2}$ reported in \cite{Heersnik}. The calculations in
\cite{Heersnik} use quantum noise operators and do not account for
the finite relaxation time of the Kerr nonlinearity. However, the
relaxation time, which was accounted here, is of a fundamental
importance since it determines the level of quantum fluctuations
of Stokes operators below the level corresponding to the coherent
state [see Eq. (\ref{S2Spectra})]. Besides, we indicated the
optimal strategy for the successful generation of the PS state in
the electronic Kerr medium.
\section{Conclusion}\label{conclusion}
We investigated the formation of polarization-squeezed light in a
nonlinear medium with electronic Kerr nonlinearity. The
correlation functions and corresponding spectra of quantum Stokes
parameters $\hat{S}_{2}$ and $\hat{S}_{3}$ were considered. We
shown that, by adjusting the linear phase difference between
pulses, the maximum suppression of the quantum fluctuations of
$\hat{S}_{2}$ or $\hat{S}_{3}$ can be realized at the spectral
component of interest. It is established that the increase of the
intensity of the control pulse can be employed to suppress the
quantum fluctuations of $\hat{S}_{2}$.  We find that the increase
of one nonlinear coefficient ($\gamma_{2}$) in comparison with
another one ($\gamma_{1}$) produces a substantial suppression of
quantum fluctuations of $\hat{S}_{2}$.

\end{document}